\documentclass[aps,prb,twocolumn,shortbibliography,superscriptaddress]{revtex4-2}
\usepackage{epsfig, epstopdf}
\usepackage{comment}
\usepackage{amsmath, amssymb, amsfonts, bm}
\usepackage{makecell, rotating, dcolumn}
\usepackage{hyperref, graphicx}
\usepackage{color, tikz, xcolor}
\usepackage[commandnameprefix=always]{changes}
\usepackage{tabularray}
\usepackage{tikz}
\usepackage{physics}
\usetikzlibrary{calc}
\usepackage[symbol]{footmisc}
\DefineFNsymbols*{noasterisk}{
   {\TextOrMath \textdagger \dagger}%
   {\TextOrMath \textdaggerdbl \ddagger}%
   {\TextOrMath \textsection  \mathsection}%
   {\TextOrMath \textparagraph \mathparagraph}%
   {\TextOrMath \textbardbl \|}%
   {\TextOrMath {\textdagger\textdagger}{\dagger\dagger}}%
   {\TextOrMath {\textdaggerdbl\textdaggerdbl}{\ddagger\ddagger}}%
}
\setfnsymbol{noasterisk}

\begin{document}
\title{Chirality-induced spin and orbital Rashba-Edelstein effects \linebreak in twisted van der Waals materials}

\author{Sebasti{\'a}n P{\'e}rez-Patricio}
\thanks{These authors contributed equally to this work.}
\affiliation{Departamento de Física, Facultad de Ciencias, Universidad Nacional Autónoma de México, Circuito Exterior s/n, Ciudad Universitaria, Coyoacán, Mexico City 04510, Mexico}

\author{Zhiren He}
\thanks{These authors contributed equally to this work.}
\affiliation{Zernike Institute for Advanced Materials, University of Groningen, Nijenborgh 3, 9747 AG Groningen, The Netherlands}

\author{Roxana Mitzay{\'e} Del Castillo V{\'a}zquez}
\affiliation{Departamento de Física, Facultad de Ciencias, Universidad Nacional Autónoma de México, Circuito Exterior s/n, Ciudad Universitaria, Coyoacán, Mexico City 04510, Mexico}

\author{Jagoda S{\l}awi{\'n}ska}
%\email{jagoda.slawinska@rug.nl}
\affiliation{Zernike Institute for Advanced Materials, University of Groningen, Nijenborgh 3, 9747 AG Groningen, The Netherlands}

\begin{abstract}
\noindent Van der Waals materials provide ways to control spin-momentum locking and charge-to-spin conversion via crystal symmetry. Using group theory analysis and Slater-Koster calculations, we show that twisted bilayers host unexpected spin textures, including symmetry-protected parallel spin-momentum locking along high-symmetry lines, independent of twist angle and constituent materials. We further demonstrate an unconventional Rashba-Edelstein effect (REE), with spin accumulation parallel to the charge current that changes sign upon reversing the twist angle, reflecting the change in chirality and resembling chirality-induced spin selectivity. The orbital REE exhibits similar magnitude and energy dependence, extending the recently proposed chirality-induced orbital selectivity to twisted materials. Density functional theory calculations for representative bilayers, including AgI and WSe$_2$, support our findings across different Bravais lattices. These results establish twisting as a means to realize chirality-dependent spin and orbital responses in van der Waals materials.
\end{abstract}

\maketitle

Spin-momentum locking couples the spin polarization of an electronic state to its momentum. It is one of the fundamental concepts in spintronics and determines a wide range of phenomena, from topological transport to spin generation via the spin Hall and Rashba-Edelstein effects (REE) \cite{Rembert2015}. In practice, this coupling is not arbitrary: it is imposed by the combination of spin-orbit coupling (SOC) and crystal symmetry, and it varies across the Brillouin zone (BZ), with different momentum-space regions constrained by distinct symmetries to host different spin-momentum locking patterns \cite{QihangLiu2024, MeraAcosta2021}. Because macroscopic responses are integrated over momenta, this variation makes it difficult to obtain charge-to-spin conversion with a desired geometry and magnitude \cite{Analogs2023, Collinear2023}.

Two-dimensional (2D) van der Waals (vdW) materials offer a different route to overcome this limitation because vdW bilayers can be tuned through stacking, sliding, and twisting \cite{Valenzuela2021}. This additional structural freedom provides a direct handle on the crystal symmetry and therefore on the spin texture itself. In graphene interfaced with transition metal dichalcogenides (TMDs), various spin textures have been reported, including conventional and tilted Rashba, Weyl-like, and Zeeman-like spin polarizations \cite{Fabian2023, Fabian2024, Przybysz2026}. This diversity is reflected in spin responses with different geometries, including conventional and unconventional REE \cite{Lee2022, Felix2024}. However, these realizations often depend strongly on the precise twist angle or on a specific substrate-induced mechanism, which limits reproducibility and makes device realization difficult.

Among different spin textures, persistent spin textures (PSTs) are of particular interest due to the potential to enhance spin lifetimes \cite{Ghiasi2017, Benitez2018, Barts2025}. Unlike Rashba or Weyl spin textures, PSTs remain unidirectional over an extended region of momentum space and can protect spin polarization against momentum-dependent spin dephasing \cite{Tsymbal2018, Kilic2025}. In 2D materials, symmetry-protected PSTs are typically associated with mirror planes, which can enforce a unidirectional spin texture across the full BZ \cite{Ji2022}. Because twisting generally breaks mirror symmetries and makes the crystal chiral \cite{Stroppa2021, APL2023}, PST has rarely been considered as a spin texture that can emerge from twisting of van der Waals materials. 

Here, we show that general twisted homobilayers host symmetry-protected PSTs along high-symmetry lines in the BZ, independently of the twist angle or constituent materials. Using group theory analysis and Slater-Koster calculations, we identify these line-like PSTs across different Bravais lattices, providing a general way to realize a PST in vdW materials without relying on Zeeman-like spin textures found in some TMDs \cite{JHGarcia2020, Bhowal2020, Jafari2024}. Consistent with the parallel spin-momentum locking along high-symmetry lines, we also demonstrate an unconventional REE, with spin accumulation parallel to the charge current \cite{Analogs2023, Collinear2023}, similar to that reported in chiral tellurium \cite{Furukawa2017, Calavalle2022, Roy2022}, but realized here in 2D vdW materials. The response reverses sign upon reversing the twist angle, reflecting the opposite chiralities of the two structures, akin to chirality-induced spin selectivity (CISS). We also calculate the orbital REE within the atom-center approximation, finding magnitudes and energy dependence similar to the spin response. This extends the concept of chirality-induced orbital selectivity (CIOS), previously proposed in other systems \cite{Yoda2015,Yoda2018, Gobel2025_1D, Gobel2025}, to 2D materials. 

We support these findings with density functional theory (DFT) calculations for twisted homobilayers with different Bravais lattices, including semiconducting AgI and WSe$_2$, as well as graphene and InBi. Together with the symmetry analysis and Slater-Koster calculations, these results demonstrate that the mechanism is not restricted to a specific material, but arises from the symmetry of bilayers. More broadly, our results extend chirality-induced spin and orbital selectivity to twisted vdW materials.

\section*{Results}
\subsection*{Symmetry analysis of twisted bilayers}

%This is because the symmetry of a $k$ point imposes constraints on the spin polarization of the Bloch states $\psi_i$. In general, for any\\

% The symmetry of a twisted bilayer is determined by both the symmetry of the individual layers and their relative orientation. Depending on the lattice geometry and the surviving symmetry operations after twisting, the resulting bilayer can belong to different space and layer groups. Monoclinic, orthorombic, tetragonal, trigonal, and hexagonal lattices are considered, for which the relevant space and layer groups, together within the corresponding PST point and lines, are summarized in Table \ref{Table1} \cite{Kilic2025}.  Furthermore, the associated Rashba-Edelstein tensor for each space group is also shown in the same table \cite{Analogs2023}. For all the considered groups, the off-diagonal components are zero, while for tetragonal, trigonal, and hexagonal lattices, $\chi_{11}=\chi_{22}$.

% \begin{itemize}
% \item Symmetry of twisted bilayer and possible space (layer) groups

% \item Symmetry enforced PST, show the requirement of in-plane $C_2$ axis for the PST lines
% \item General strategy to create/preserve in-plane $C_2$ axis with flipped top layer
% \end{itemize}

%From a symmetry perspective, we demonstrate how unconventional REE, where the induced magnetization is parallel to the electric field, arises in \emph{all} twist homobilayer, and switches sign when structural chirality is reversed. We also formulate the symmetry basis i.e. in-plane $C_2$ and $2_1$ screw axis, that enables PST at high-symmetry lines and points.

Crystal symmetry determines both the spin textures and the macroscopic responses of twisted bilayers. We therefore begin with a general analysis to identify the symmetries of these systems, the emergence of persistent spin textures, and the consequences for the REE.

The symmetry of a twisted bilayer can be determined from the symmetry operations that survive twisting, namely those that either make both top and bottom layer invariant, or exchange the bilayer. As detailed in Methods, all twisted bilayers are necessarily chiral, lacking inversion and all mirror planes, unless they are accidentally restored at special angles, while the proper rotations are preserved. The in-plane $C_{2}$ axis is of particular interest, because it exchanges the top and bottom layer. Even if the group of the monolayer does not contain in-plane $C_{2}$ rotation, flipping one of the layers along an in-plane axis by 180$^\circ$ will artificially create them, making the top and bottom layer related by one or more in-plane $C_{2}$ symmetries \cite{Liu2024,Pan2024}. For example, in a monolayer with layer group (LG) p4/m, a direct twist will lead to LG p4. The flip operation adds 4 $C_2$ rotations $C_{2[10]}$, $C_{2[01]}$, $C_{2[11]}$, $C_{2[1\bar{1}]}$, leading to LG p422. 

The 2$_1$ screw axes, which occur in square, rectangular and oblique monolayers, can survive the twist but only in certain commensurate moir\'e cells. We illustrate this using LG p4/nmm (64), to which a large number of monolayers belong. A direct twist without flipping preserves $C_{2[11]}$, $C_{2[1\bar{1}]}$ as well as screw axes $\{C_{2[10]}|1/2,1/2\}$, $\{C_{2[01]}|1/2,1/2\}$ when the commensurate moir\'e supercell indices ($n,m$)=(2,1),(4,1)$\dots$, i.e. $n-m$ is odd. This gives LG p42$_1$2 (54). Otherwise, 2$_1$ screw axes become $C_2$ axes, leading to LG p422 (53). For rectangular and oblique lattices, the twisted lattices are in general incommensurate and screw axis is unlikely to be preserved in real twisted bilayers. Nevertheless, we include the layer groups with screw axis in our analysis for completeness.

These symmetries are important because they can enforce PSTs. In particular, in-plane $C_2$ symmetries can restrict the spin expectation values along high-symmetry lines of the BZ, while in-plane 2$_1$ screw axes can impose analogous constraints at high-symmetry points, providing a symmetry-based mechanism for the emergence of PSTs \cite{Kilic2025}. As shown in Methods and Table \ref{tab:pst}, both symmetries would enforce the spin direction to be parallel or anti-parallel to the rotation/screw axis. For example, in a square lattice with $C_{2[11]}$ and $C_{2[1\bar{1}]}$, the spins on the line from $\Gamma$(0,0) to M($\pm$1/2,$\pm$1/2) must be aligned along the diagonal of the BZ. 

%Since the $C_2$ axes forbid out-of-plane electric polarization
Following previous arguments, twisted homobilayers are chiral and in-plane $C_2$ axes can be created by flipping. Consequently, they will show a purely collinear REE, where accumulated spins $S_i$ are aligned with the applied electric field $E_j$ with $i=j$ (see Table \ref{tab:pst}). The collinear Rashba-Edelstein response $\chi_{ii}$, defined by $S_i=\chi_{ii}E_i$ has the opposite sign for mirror-related twisted bilayers with opposite chirality, i.e. twist angles $\theta$ and $-\theta$. This is because electric field $E_i$ is a polar vector, while spin accumulation $S_i$ is an axial (or pseudo) vector that acquires an additional minus sign under an improper spatial operation such as a mirror reflection. 

This can be also interpreted in terms of the spin textures. The symmetry-allowed effective SOC gives rise to collinear spin-momentum locking that takes the form of $\vec{k}\cdot\vec{s}$, a pseudoscalar that transforms as chirality. The resulting spin accumulation arises from the electric-field-induced nonequilibrium redistribution of occupations of states with momentum-dependent spin polarization near the Fermi surface, and therefore inherits this sign reversal. In addition, as orbital magnetization behaves the same as spin magnetization in terms of symmetry, our analysis extends naturally to orbital REE, even though the orbital REE and orbital textures originate from orbital hybridization and do not require SOC \cite{Go2021,Cysne2025}.

Although similar in terms of symmetry, and hence macroscopic responses, chiral twisted bilayers can have very different spin textures from chiral 3D bulk crystals \cite{Martinez2022}. The in-plane $C_2$ axis that exchanges top and bottom layer also divide the BZ into regions related by $C_2$ symmetry, as shown in the schematic images in Table \ref{tab:pst}. Without interlayer interaction, the alternating gray and white region will contain purely states from top and bottom layer. With weak interlayer interaction, each region will still be top/bottom layer dominant, as shown by an earlier study in twisted bilayer graphene \cite{Stroppa2021}. As the top and bottom layer are related by in-plane $C_2$, each layer locally would only retain $C_{nz}$ symmetry, allowing dipoles along $\hat{z}$ that are anti-aligned between the two layers. This can also be understood by the vertical asymmetry in electrostatic potential, with a monolayer on one side and vacuum on the other. As a consequence, there are regions in BZ zone away from the $C_2$ axis that can \emph{locally} show a dominantly Rashba spin texture $\vec{k}\times\vec{s}$ in addition to the collinear $\vec{k}\cdot\vec{s}$, while in general, terms higher-order in $\vec{k}$ can also play a role. 

In Table \ref{tab:pst}, we list all possible layer groups of a twisted homobilayer and where symmetry-enforced PSTs are located. The layer groups are classified based on their lattice type, each with a corresponding BZ and form of the REE response tensor.

%In 3D, local electric polarization develops only for certain Wyckoff sites. For example, in space group P622, the Wyckoff site 2e at (0,0,z) and (0,0,-z) with site symmetry 6 can support opposite local electric dipoles, but not at Wyckoff site 2c at (1/3,2/3,0) and (2/3,1/3,0) with site symmetry 32. 

\newcommand{\SquareBZ}{
	\begin{tikzpicture}[scale=1.8]
		
		% Corner points of the square BZ
		\coordinate (A) at ( 0.5, 0.5);
		\coordinate (B) at ( 0.0, 0.5);
		\coordinate (C) at (-0.5, 0.5);
		\coordinate (D) at (-0.5, 0.0);
		\coordinate (E) at (-0.5,-0.5);
		\coordinate (F) at ( 0.0,-0.5);
		\coordinate (G) at ( 0.5,-0.5);
		\coordinate (H) at ( 0.5, 0.0);
		\coordinate (O) at ( 0.0, 0.0);
		
		% Fill the 8 sectors with alternating colors
		\fill[gray, opacity=0.5] (O) -- (A) -- (B) -- cycle;
		\fill[gray, opacity=0.5] (O) -- (C) -- (D) -- cycle;
		\fill[gray, opacity=0.5] (O) -- (E) -- (F) -- cycle;
		\fill[gray, opacity=0.5] (O) -- (G) -- (H) -- cycle;
		
		% Outer square boundary
		\draw[thick] (-0.5,-0.5) rectangle (0.5,0.5);
		
		% dotted lines
		\draw[dotted, thick] (O) -- (A);
		\draw[dotted, thick] (O) -- (B);
		\draw[dotted, thick] (O) -- (C);
		\draw[dotted, thick] (O) -- (D);
		\draw[dotted, thick] (O) -- (E);
		\draw[dotted, thick] (O) -- (F);
		\draw[dotted, thick] (O) -- (G);
		\draw[dotted, thick] (O) -- (H);
		
		% Labels
		\node[] at (O) {$\Gamma$};
		\node[right]      at (H) {$Y$};
		\node[above right] at (A) {$M$};
		\node[above]       at (B) {$X$};
		
	\end{tikzpicture}
}
\newcommand{\RecBZ}{
	\begin{tikzpicture}[scale=1.8]
		
		% Corner points
		\coordinate (A) at ( 0.6, 0.4);
		\coordinate (B) at ( 0.0, 0.4);
		\coordinate (C) at (-0.6, 0.4);
		\coordinate (D) at (-0.6, 0.0);
		\coordinate (E) at (-0.6,-0.4);
		\coordinate (F) at ( 0.0,-0.4);
		\coordinate (G) at ( 0.6,-0.4);
		\coordinate (H) at ( 0.6, 0.0);
		\coordinate (O) at ( 0.0, 0.0);
		
		% Fill the sectors with alternating colors
		\fill[gray, opacity=0.5] (O) -- (B) -- (C) -- (D)-- cycle;
		\fill[gray, opacity=0.5] (O) -- (F) -- (G) -- (H) -- cycle;
		
		% Outer boundary
		\draw[thick] (-0.6,-0.4) rectangle (0.6,0.4);
		
		% lines
		\draw[dotted, thick] (O) -- (H);
		\draw[dotted, thick] (O) -- (B);
		\draw[dotted, thick] (O) -- (D);
		\draw[dotted, thick] (O) -- (F);
		
		% Labels
		\node[] at (O) {$\Gamma$};
		\node[right]      at (H) {$X$};
		\node[above right] at (A) {$S$};
		\node[above]       at (B) {$Y$};
		
	\end{tikzpicture}
}

\newcommand{\ObliqueBZ}{
	\begin{tikzpicture}[scale=1.8]
		
		% Corner points
		\coordinate (A) at ( 0.6, 0.4);
		\coordinate (B) at ( 0.0, 0.4);
		\coordinate (C) at (-0.6, 0.4);
		\coordinate (D) at (-0.6, 0.0);
		\coordinate (E) at (-0.6,-0.4);
		\coordinate (F) at ( 0.0,-0.4);
		\coordinate (G) at ( 0.6,-0.4);
		\coordinate (H) at ( 0.6, 0.0);
		\coordinate (O) at ( 0.0, 0.0);
		
		% Fill the sectors with alternating colors
		\fill[gray, opacity=0.5] (C) -- (E) -- (F) -- (B)-- cycle;
		
		% Outer boundary
		\draw[thick] (-0.6,-0.4) rectangle (0.6,0.4);
		
		% lines
		\draw[dotted, thick] (O) -- (B);
		\draw[dotted, thick] (O) -- (F);
		
		% Labels
		\node[] at (O) {$\Gamma$};
		\node[right]      at (H) {$Y$};
		\node[above right] at (A) {$C$};
		\node[above]       at (B) {$Z$};
		
	\end{tikzpicture}
}

\newcommand{\HexBZ}{
	\begin{tikzpicture}[scale=1,rotate=30]
		
		% Center
		\coordinate (G) at (0,0);
		
		% Hexagon vertices
		\coordinate (K1) at ({cos(30)}, {sin(30)});
		\coordinate (K2) at (0,  1);
		\coordinate (K3) at ({-cos(30)}, {sin(30)});
		\coordinate (K4) at ({-cos(30)}, {-sin(30)});
		\coordinate (K5) at (0, -1);
		\coordinate (K6) at ({cos(30)}, {-sin(30)});
		
		% Midpoints of edges (M points)
		\coordinate (M1) at ($(K6)!0.5!(K1)$); 
		\coordinate (M2) at ($(K1)!0.5!(K2)$);
		\coordinate (M3) at ($(K2)!0.5!(K3)$);
		\coordinate (M4) at ($(K3)!0.5!(K4)$);
		\coordinate (M5) at ($(K4)!0.5!(K5)$);
		\coordinate (M6) at ($(K5)!0.5!(K6)$);
		
		% Fill the 12 sectors with alternating colors
		\fill[gray, opacity=0.5]  (G) -- (M1) -- (K6) -- cycle;
		\fill[gray, opacity=0.5]  (G) -- (M6) -- (K5) -- cycle;
		\fill[gray, opacity=0.5]  (G) -- (M5) -- (K4) -- cycle;
		\fill[gray, opacity=0.5]  (G) -- (M4) -- (K3) -- cycle;
		\fill[gray, opacity=0.5]  (G) -- (M3) -- (K2) -- cycle;
		\fill[gray, opacity=0.5]  (G) -- (M2) -- (K1) -- cycle;
		
		% Outer hexagon boundary
		\draw[thick] (K1) -- (K2) -- (K3) -- (K4) -- (K5) -- (K6) -- cycle;
		
		% Dotted lines
		\draw[dotted, thick] (G) -- (M1);
		\draw[dotted, thick] (G) -- (K1);
		\draw[dotted, thick] (G) -- (M2);
		\draw[dotted, thick] (G) -- (K2);
		\draw[dotted, thick] (G) -- (M3);
		\draw[dotted, thick] (G) -- (K3);
		\draw[dotted, thick] (G) -- (M4);
		\draw[dotted, thick] (G) -- (K4);
		\draw[dotted, thick] (G) -- (M5);
		\draw[dotted, thick] (G) -- (K5);
		\draw[dotted, thick] (G) -- (M6);
		\draw[dotted, thick] (G) -- (K6);
		
		% Labels
		\node[]  at (G)  {$\Gamma$};
		\node[right]       at (M1) {$M$};
		\node[above right] at (K1) {$K$};
		
	\end{tikzpicture}
}
\newcommand{\TriBZ}{
	\begin{tikzpicture}[scale=1,rotate=30]
		
		% Center
		\coordinate (G) at (0,0);
		
		% Hexagon vertices
		\coordinate (K1) at ({cos(30)}, {sin(30)});
		\coordinate (K2) at (0,  1);
		\coordinate (K3) at ({-cos(30)}, {sin(30)});
		\coordinate (K4) at ({-cos(30)}, {-sin(30)});
		\coordinate (K5) at (0, -1);
		\coordinate (K6) at ({cos(30)}, {-sin(30)});
		
		% Midpoints of edges (M points)
		\coordinate (M1) at ($(K6)!0.5!(K1)$);
		\coordinate (M2) at ($(K1)!0.5!(K2)$);
		\coordinate (M3) at ($(K2)!0.5!(K3)$);
		\coordinate (M4) at ($(K3)!0.5!(K4)$);
		\coordinate (M5) at ($(K4)!0.5!(K5)$);
		\coordinate (M6) at ($(K5)!0.5!(K6)$);
		
		% Fill the sectors with alternating colors
		\fill[gray, opacity=0.5]  (G) -- (K6) -- (K5) -- cycle;
		\fill[gray, opacity=0.5]  (G) -- (K4) -- (K3) -- cycle;
		\fill[gray, opacity=0.5]  (G) -- (K2) -- (K1) -- cycle;
		
		% Outer hexagon boundary
		\draw[thick] (K1) -- (K2) -- (K3) -- (K4) -- (K5) -- (K6) -- cycle;
		
		% Dotted lines
		\draw[dotted, thick] (G) -- (K1);
		\draw[dotted, thick] (G) -- (K2);
		\draw[dotted, thick] (G) -- (K3);
		\draw[dotted, thick] (G) -- (K4);
		\draw[dotted, thick] (G) -- (K5);
		\draw[dotted, thick] (G) -- (K6);
		
		% Labels
		\node[]  at (G)  {$\Gamma$};
		\node[right]       at (M1) {$M$};
		\node[above right] at (K1) {$K$};
		
	\end{tikzpicture}
}
\newcommand{\RotTriBZ}{
	\begin{tikzpicture}[scale=1,rotate=30]
		
		% Center
		\coordinate (G) at (0,0);
		
		% Hexagon vertices
		\coordinate (K1) at ({cos(30)}, {sin(30)});
		\coordinate (K2) at (0,  1);
		\coordinate (K3) at ({-cos(30)}, {sin(30)});
		\coordinate (K4) at ({-cos(30)}, {-sin(30)});
		\coordinate (K5) at (0, -1);
		\coordinate (K6) at ({cos(30)}, {-sin(30)});
		
		% Midpoints of edges (M points)
		\coordinate (M1) at ($(K6)!0.5!(K1)$);
		\coordinate (M2) at ($(K1)!0.5!(K2)$);
		\coordinate (M3) at ($(K2)!0.5!(K3)$);
		\coordinate (M4) at ($(K3)!0.5!(K4)$);
		\coordinate (M5) at ($(K4)!0.5!(K5)$);
		\coordinate (M6) at ($(K5)!0.5!(K6)$);
		
		% Fill the sectors with alternating colors
		\fill[gray, opacity=0.5]  (G) -- (M2) -- (K2) -- (M3) -- cycle;
		\fill[gray, opacity=0.5]  (G) -- (M4) -- (K4) -- (M5) -- cycle;
		\fill[gray, opacity=0.5]  (G) -- (M6) -- (K6) -- (M1) -- cycle;
		
		% Outer hexagon boundary
		\draw[thick] (K1) -- (K2) -- (K3) -- (K4) -- (K5) -- (K6) -- cycle;
		
		% Dotted lines
		\draw[dotted, thick] (G) -- (M1);
		\draw[dotted, thick] (G) -- (M2);
		\draw[dotted, thick] (G) -- (M3);
		\draw[dotted, thick] (G) -- (M4);
		\draw[dotted, thick] (G) -- (M5);
		\draw[dotted, thick] (G) -- (M6);
		
		% Labels
		\node[]  at (G)  {$\Gamma$};
		\node[right]       at (M1) {$M$};
		\node[above right] at (K1) {$K$};
		
	\end{tikzpicture}
}

\begin{table*}[!ht]
	\caption{List of all possible layer groups of a twisted homobilayer, excluding LG p4 (49), p3 (65) and p6 (73), which do not have PST anywhere, and can have finite REE component $\chi_{12}=-\chi_{21}$ from out-of-plane polarization. For each lattice type, we sketch the Brillouin zone, which is divided into grey and white regions related by $C_2$ axes (dashed lines). For each layer group, we show regions where PST can occur. Following the classification in Ref. \cite{Kilic2025}, type I PSTs for nondegenerate states (normal font) are located along high-symmetry lines, while type II PSTs for degenerate state (bold font) can occur at high-symmetry points. The spin direction is indicated in square brackets and expressed in terms of the conventional reciprocal lattice vectors. \label{tab:pst}}
	\begin{tblr}{
			colspec = {cccccc},
            vspan = even,
			hline{1,Z} = {1}{-}{solid},
			hline{1,Z} = {2}{-}{solid},
			cell{2}{1-3} = {r=2}{c},
			cell{4}{1-3} = {r=6}{c},
			cell{10}{1-3} = {r=5}{c},
			cell{15}{1-3} = {r=2}{c,m},
            cell{17}{1-6} = {r=2}{c,m}
		}
		Lattice & Brillouin zone & REE & Layer group &  PST points & PST lines \\ \hline
		Oblique  & \ObliqueBZ & $\begin{bmatrix}
				\chi_{11} & 0 \\ 0 & \chi_{22}
			\end{bmatrix}$  & p112$_y$ (3), p2$_y$11 (8) &  -- & [01] $\Gamma Z$ \\ \cline[dashed]{3-7}
		& & & p2$_{1y}$11 (9) & [01] $\bm{Z}$ & [01] $\Gamma Z$ \\ \hline
		Rectangular & \RecBZ& $\begin{bmatrix}
				\chi_{11} & 0 \\ 0 & \chi_{22}
			\end{bmatrix}$ & p222 (19) & -- & [10] $\Gamma X$, $YS$ \\
		& & & &  & [01] $\Gamma Y$, $XS$ \\ \cline[dashed]{3-7}
		& &  & p2$_{1y}$22 (20) & [01] $\bm{Y}$ & [10] $\Gamma X$, $YS$\\
		& & & &  & [01] $\Gamma Y$\\ \cline[dashed]{3-7}
		& & & p2$_1$2$_1$2 (21) & [10] $\bm{X}$ & [10] $\Gamma X$ \\
		& & & &   [01] $\bm{Y}$ & [01] $\Gamma Y$\\ \hline
		Square & \SquareBZ & $\begin{bmatrix}
				\chi_{11} & 0 \\ 0 & \chi_{11}
			\end{bmatrix}$&  p422 (53) & -- & [01] $\Gamma X$ \\
		& & & &  & [11] $\Gamma M$ \\
		& & & &  & [10] $XM$\\ \cline[dashed]{3-7}
		& & & p42$_{1y}$2 (54) & [01] $\bm{X}$ & [01] $\Gamma X$ \\
		& & & &  & [11] $\Gamma M$  \\ \hline
		   Triangular & \RotTriBZ \TriBZ  & $\begin{bmatrix}
				\chi_{11} & 0 \\ 0 & \chi_{11}
			\end{bmatrix}$  &  p312 (67) &  -- & [10] $\Gamma M$ \\ \cline[dashed]{3-7}
		& & & p321 (68) &  -- & [11] $\Gamma K$ \\ \hline
		Hexagonal & \HexBZ & $\begin{bmatrix}
				\chi_{11} & 0 \\ 0 & \chi_{11}
			\end{bmatrix}$ &  p622 (76) & -- & [11] $\Gamma K$, [10] $\Gamma M$ \\
        & & & & & \\
	\end{tblr}
\end{table*}

\begin{figure*}
    \centering
    \includegraphics[width=1.0\linewidth]{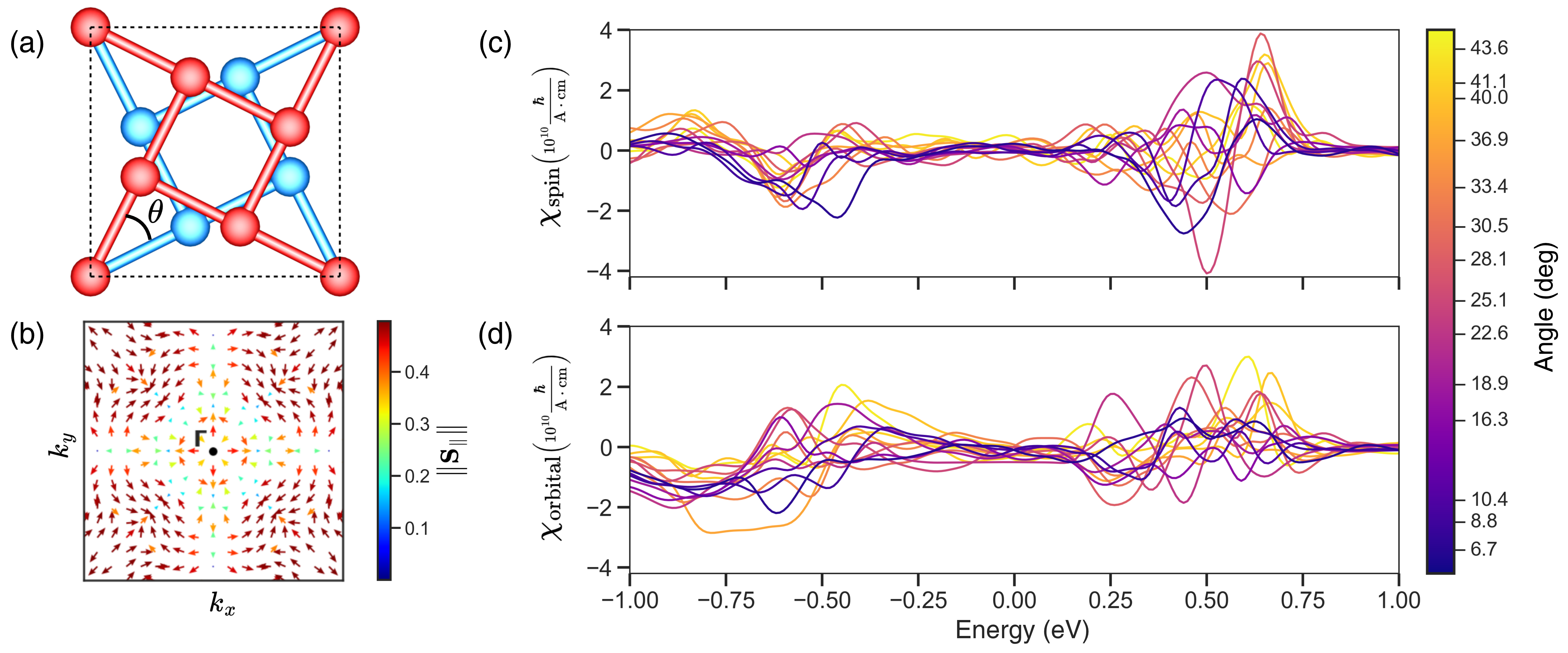}   
    \caption{(a) Schematic representation of a twisted square-lattice bilayer, where the upper layer (red) is rotated by a twist angle $\theta$ relative to the lower layer (blue). (b) Spin texture obtained from Slater-Koster calculations, mapped over the BZ region $[-0.20,\:0.20]\times[-0.20,\:0.20]$ in fractional reciprocal coordinates, centered at $\Gamma$; the color scale indicates the in-plane spin magnitude $\|\mathbf{S}_{\parallel}\|$. (c) Spin and (d) orbital Rashba-Edelstein effects calculated using the Slater-Koster Hamiltonians for various commensurate twist angles, distinguished by color. The components $\chi_{xx}$ and $\chi_{yy}$ are equal by symmetry.}
    \label{fig:ekai_tb}
\end{figure*}

\subsection*{Slater-Koster calculations of general twisted bilayers}
The spin textures and charge-to-spin conversion are further studied using the Slater-Koster (SK) tight-binding method, which parameterizes orbital-dependent hopping matrix elements based on the relative atomic geometry \cite{SK1954}. Such calculations for general bilayers offer two key advantages. First, a relatively small orbital basis makes it computationally feasible to study a wide range of commensurate twist angles, including those with large moir\'e supercells. This allows us to systematically assess the twist-angle dependence of REE, which is of critical importance to device applications where precise and reproducible control of the twist angle remains challenging. Second, the model parameters can be tuned to analyze the effects of interlayer interactions and SOC strength on the REE response, which is difficult to achieve at the level of first-principles calculations. This helps us assess the robustness of the predicted responses and ensure that our observations are not tied to particular parameters.

Figure \ref{fig:ekai_tb}a illustrates one of the considered square lattices, in which two monolayers, each containing one atomic site with $p_x$, $p_y$, and $p_z$ orbitals per unit cell, are rotated to form a twisted homobilayer described by LG p422 (53). The SK Hamiltonian is constructed using the hopping parameters specified in Methods. The representative spin textures centered at $\Gamma$ are shown in Fig. \ref{fig:ekai_tb}b, revealing PSTs along the horizontal, vertical, and diagonal directions, in agreement with the symmetry analysis. Figure \ref{fig:ekai_tb}c and d show the spin and orbital REE, respectively, as a function of Fermi level for several commensurate twist angles ranging from 6.7$^\circ$ to 43.6$^\circ$. Although both responses can vary with twist angle at a particular energy, the order of magnitude remains the same across considered angles, and their dependence on chemical potential changes slowly with the twist angle.

To verify the robustness of these results, we perform the calculations with tuned model parameters. Although the resulting spin and orbital REE responses change, they exhibit similar magnitudes across all selected twist angles. Reducing the SOC strength by a factor of 10 significantly suppresses the spin REE, while the orbital REE is only slightly affected (see Supplementary Material, Fig. S1). This corroborates with the distinct origin of spin and orbital REE. Reducing the interlayer interactions suppresses both spin and orbital REE responses (see Fig. S2). Such behavior is expected because twisting modifies the interlayer orbital hybridization, which gives rise to the momentum-dependent spin and orbital textures associated with structural chirality. In the limit of vanishing interlayer interactions, the individual layers remain centrosymmetric and exhibit no spin splitting. 

\subsection*{First-principles calculations for representative materials}
To illustrate the predicted spin textures and Rashba-Edelstein effect in realistic homobilayers, we consider two representative vdW materials: semiconducting AgI and WSe$_2$, with square and hexagonal lattices, respectively. We also present calculations for twisted bilayer graphene and metallic InBi in the Supplementary Material.
\vskip 0.3 em

\begin{figure*}
    \centering
    \includegraphics[width=1.0\linewidth]{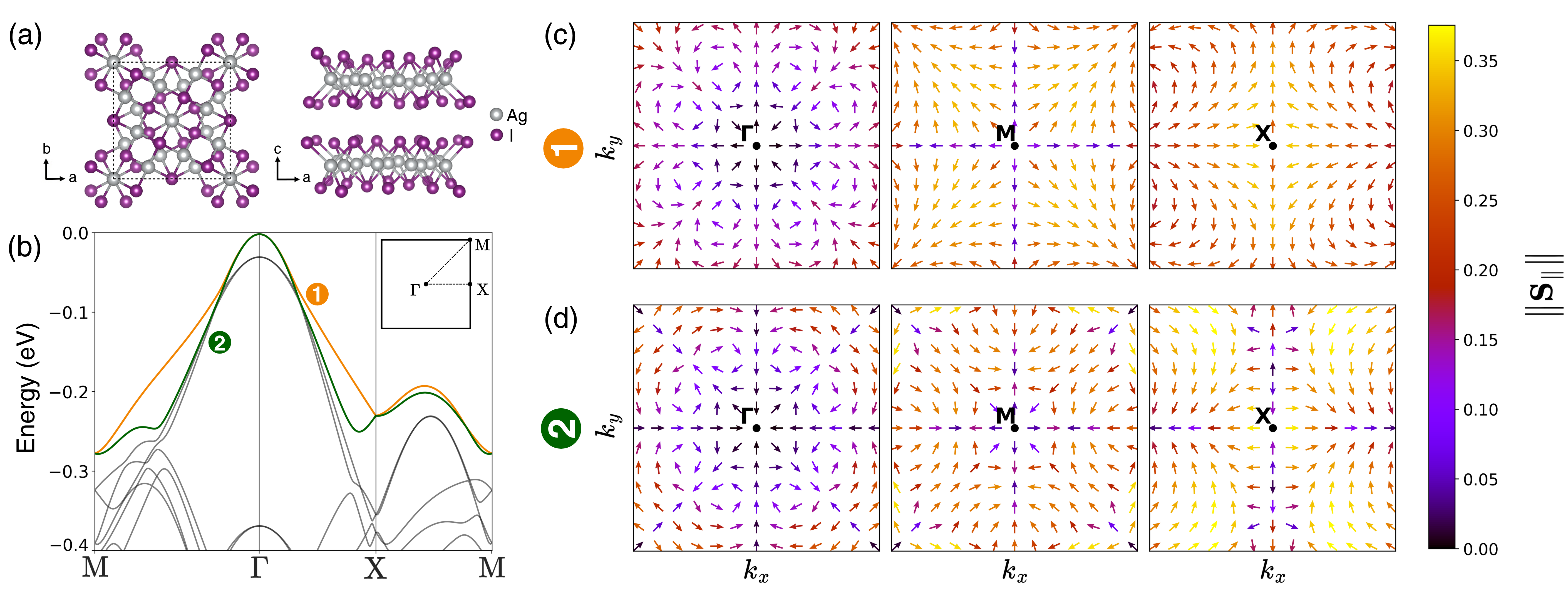}
    \caption{(a) Top (left) and side (right) views of the tb-AgI structure. (b) Calculated band structure with the VBM set at zero; the first and second topmost valence bands are shown in color and labeled. (c) Spin textures around the $\Gamma$, $M$, and $X$ points for the first band. (d) Same as (c), but for the second band. All spin textures are plotted over a square region $[-0.25,\:0.25]\times[-0.25,\:0.25]$ in fractional reciprocal coordinates, centered at each high-symmetry point in the BZ. The arrow color indicates the in-plane magnitude of the spin expectation value $\|\mathbf{S}_{\parallel}\|$, while the out-of-plane component is zero by symmetry.}
    \label{fig:fig1}
\end{figure*}

\textbf{Twisted bilayer silver iodide (tb-AgI).} Silver iodide (AgI) is a semiconductor that exhibits several structural phases in bulk form  \cite{ExpAgI2025, TheoAgI2022}. Among them, a tetragonal layered phase (P4/nmm) has been synthesized experimentally \cite{Moore1968}, providing a suitable starting point for exploring AgI in the 2D limit. Theoretically exfoliated monolayers derived from this phase have also been reported to be stable and semiconducting \cite{Mounet2018, C2DB_1, C2DB_2}. Starting from a square monolayer described by LG p4/nmm (64), we construct a commensurate twisted bilayer AgI structure (tb-AgI) using a ($\sqrt{5}\times\sqrt{5}$) supercell and a twist angle of $\theta=36.87^\circ$. The top layer is 180$^\circ$ flipped along an in-plane axis, which reduces the $2_1$ screws to 2-fold rotations, and the resulting 40-atom moir\'e square superlattice, shown in Fig. \ref{fig:fig1}a belongs to LG p422 (53).

The twisted bilayer is semiconducting, with a direct band gap of $1.56$ eV at the $\Gamma$ point. This value is comparable to the gap of 1.87 eV reported for the square monolayer using first-principles calculations within the Perdew-Burke-Ernzerhof (PBE) approximation \cite{C2DB_1, C2DB_2}. The full band structure is shown in the Supplementary Material (Fig. S3); here, we focus on the valence bands (Fig. \ref{fig:fig1}b), since doping is required to access the charge-to-spin conversion effects. The spin splitting is negligible near the valence band maximum (VBM), but increases away from $\Gamma$, exceeding 50 meV near the $X$ point.

The spin textures of the first and second valence bands are presented in Fig. \ref{fig:fig1}c-d, respectively. The nearly opposite spin textures of them reflect the fact that they are a spin-split pair. Since the system exhibits a combined $C_{2z}$ and time-reversal symmetry $\mathcal{T}$, the out-of-plane component of the spin vanishes throughout the entire BZ. We note that $C_{2z}\mathcal{T}$ leaves $(k_x,k_y)$ invariant while reversing the spin, such that $(C_{2z}\mathcal{T})S_z(k_x,k_y)=-S_z(k_x,k_y)$, thus $S_z(k_x,k_y)=0$. Consequently, the spin textures are entirely determined by their in-plane components, and the negligible numerical values of $S_z$ are not shown. 

To analyze the spin textures, we focus on the high-symmetry points. Around $\Gamma$ and $M$, both bands exhibit a four-fold symmetric pattern, consistent with the $D_4$ little group associated with these points. This symmetry gives rise to eight symmetrically related regions distributed across the four quadrants. In contrast, the $D_2$ little group of the $X$ point imposes only two-fold symmetries on the spin textures. The most important feature of these textures is the presence of PSTs along the $[10]$, $[01]$, and $[11]$ directions, in agreement with our general analysis summarized in Table \ref{tab:pst}. These line-like PSTs are clearly visible around the $\Gamma$ and $M$ points, whereas at $X$ point they occur along principal axes but not along the diagonals. The difference reflects the distinct little group that describes the $X$ point. Although the spin direction remains well defined along the PST lines, it reverses in some cases. For example, along $[11]$ PSTs of the second band, the spin reverses upon moving away from $\Gamma$, possibly due to the hybridization with deeper valence bands. 
%However, all along the $[10]$ and $[11]$ directions, line-like PSTs for the first band keep the same orientation all along the k-space, while for the second band, $[10]$ is the direction that remains PSTs unidireccional.

% Away from the high-symmetry lines, a flower-like shape centered at $\Gamma$ can be seen in Fig. \ref{fig:fig1}c-d, where the points $M$ and $X$ define the limits of this shape. The petals seem to rotate around fixed points near $\Gamma$, with the rotation occurring from one PST line to another. Furthermore, it can be seen that the ST in between these symmetry lines rotate in the opposite direction from one region in one band to the same region in the other band; this can be better seen around the $\Gamma$ point \textcolor{blue}{Not clear what rotation means}. The nearly opposite ST of the first and second valence bands reflect the fact that they are a spin-split pair.

Although local spin textures do not fully determine the REE, they provide insight into the possible spin responses. The spin textures in Fig. \ref{fig:fig1}c-d do not fit the conventional tangential or radial prototypes, but are still consistent with the symmetry-allowed form of the response tensor, for which only the diagonal components are nonzero, with $\chi_{xx}=\chi_{yy}$ and $\chi_{xy}=\chi_{yx}=0$ (see Table \ref{tab:pst}). Such diagonal components are typically associated with parallel spin-momentum locking, which is present here along the high-symmetry lines. Yet, the total response depends on contributions from the entire BZ.

\begin{figure}[h!]
    \centering
   \includegraphics[width=0.9\linewidth]{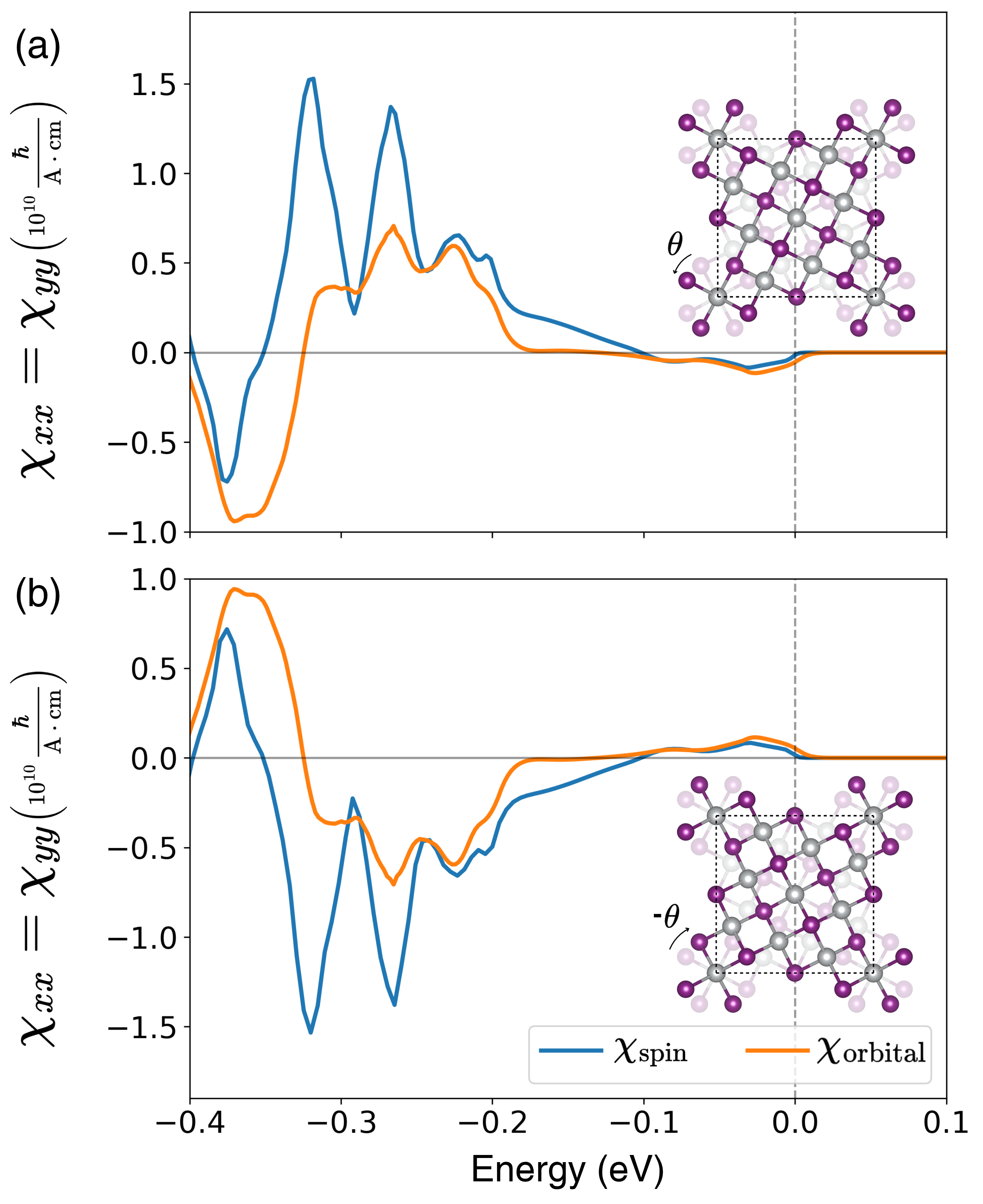}   
\caption{Spin and orbital Rashba-Edelstein effects as a function of the chemical potential calculated for the tb-AgI system at (a) a positive twist angle $\theta=36.87^\circ$ (counterclockwise rotation) and (b) a negative twist angle $\theta=-36.87^\circ$.}
    \label{fig:fig2}
\end{figure}

Figure \ref{fig:fig2} shows the REE calculated for two AgI homobilayers with opposite twist angles, in which the upper layer is rotated counterclockwise or clockwise relative to the lower layer, as shown in Fig. \ref{fig:fig2}a and b, respectively. The susceptibility $\chi$ reverses sign between the two structures, which have opposite chiralities. This sign reversal means that a given charge current generates spin accumulation in opposite directions for the two chiralities, with the spin accumulation parallel or antiparallel to the current. The resulting response is therefore chirality-dependent, resembling CISS in chiral molecules.  

The spin REE exhibits small values near the VBM, followed by a sign change and three pronounced peaks within the considered energy window. These peaks may be associated with the larger spin splittings observed in the band structure, although a direct correspondence is difficult to establish because the REE integrates contributions from the entire BZ. The calculated susceptibility reaches values on the order of $1.0-1.5\times10^{10}\frac{\hbar}{\text{A}\cdot \text{cm}}$, which is comparable with values reported for other materials with strong SOC \cite{Analogs2023, Collinear2023}, but realized here in a chiral vdW material. We also calculate the orbital REE for both clockwise- and counterclockwise-twisted structures. Its value closely follows that of the spin REE, particularly at lower energies, although the orbital response is generally smaller in magnitude. The orbital and spin responses exhibit similar signs and energy dependence, so their contributions will add constructively to the total angular momentum accumulation.

\begin{figure*}
    \centering
    \includegraphics[width=1.0\linewidth]{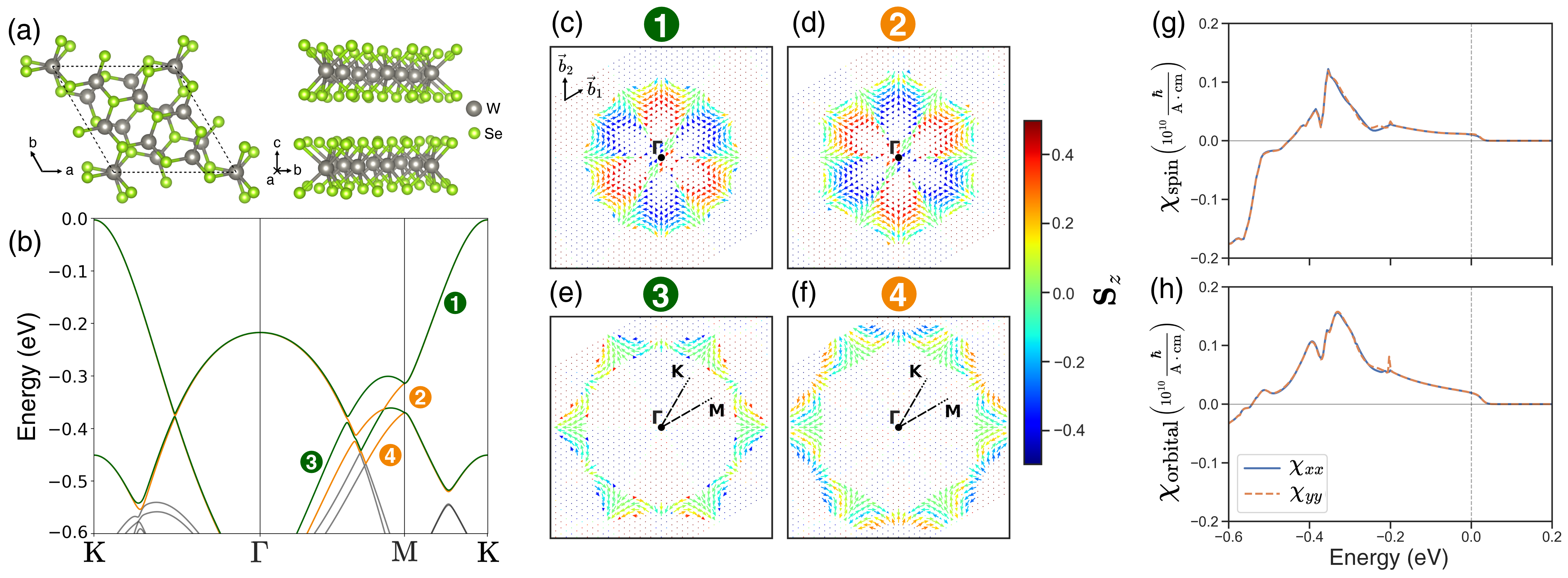}
    \caption{(a) Top (left) and side (right) views of the tb-WSe$_{2}$ structure. (b) Band structure with the VBM set at zero. Spin textures centered at $\Gamma$ for the (c) first (d) second (e) third and (f) fourth topmost valence bands, for which the color of each arrow corresponds to the out-of-plane spin expectation value $S_z$, and the length of the arrow reflects the in-plane spin polarization. All spin textures are evaluated over a square region spanning $[-\sqrt{3}b/4,\sqrt{3}b/4]\times[-\sqrt{3}b/4,\sqrt{3}b/4]$, where $b=|\vec{b}_1|=|\vec{b}_2|$ is the length of reciprocal lattice vector. The dotted lines indicate the high-symmetry paths from $\Gamma$ to $K$ and $M$. (g) Spin and (h) orbital Rashba-Edelstein effects calculated as a function of the chemical potential.}
    \label{fig:WSe2}
\end{figure*}

\vskip 0.5 em
\textbf{Twisted bilayer WSe$_2$ (tb-WSe$_{2}$).} We next consider hexagonal WSe$_2$, one of the most widely studied TMDs. The twisted bilayer system analyzed here consists of two monolayers (LG p$\bar{6}$m2 (78)) rotated relative to each other by a twist angle of $\theta = 21.9^\circ$, resulting in a ($\sqrt{7}\times\sqrt{7}$) supercell with LG p321 (68), shown in Fig. \ref{fig:WSe2}a. The electronic structure is reported in Fig \ref{fig:WSe2}b and in Fig. S3, and similarly to the bulk and monolayer, exhibits semiconducting behavior. However, the bands along the high-symmetry lines differ from those of the monolayer \cite{C2DB_1, C2DB_2}, in line with the previous studies \cite{Fabian2024}. 
%Even though the VBM is still at the $K$ point, the giant band splitting of the monolayer around this point is here absent. 

The $\Gamma$-centered spin textures of the first and second topmost valence bands are shown in Fig. \ref{fig:WSe2}c-d. They exhibit flower-like spin textures similar to AgI, although with a six-petal shape. The little group at $\Gamma$ is $D_3$, implying that the spin textures should be compatible with threefold rotational symmetry. However, the textures of these two bands exhibit a slight deviation from this symmetry near $\Gamma$, where the $C_{3z}$ operation is not exactly preserved. Interestingly, this slight symmetry-reduction becomes less pronounced farther away from $\Gamma$, where the spin textures recover a more evident threefold rotational symmetry. A possible explanation for this behavior is the near-degeneracy between these bands along the high-symmetry path $K-\Gamma-M$ (see Fig \ref{fig:WSe2}b). The interaction and mixing between degenerate or nearly degenerate states may prevent the individual band-resolved spin texture from exhibiting the expected rotational symmetry. %even though the underlying crystal symmetry remains preserved. 
%Thus, the symmetry reduction may arise from the band-resolved character of the spin textures, rather than crystal symmetry breaking.

In Fig. \ref{fig:WSe2}e-f, the spin textures for the third and fourth bands are shown; far from the $\Gamma$ point the in-plane components have notable values, and a 3-fold symmetric pattern is clearly seen, but near $\Gamma$ the in-plane magnitudes are null, for which the out-of-plane magnitude ${S}_z$ takes all the spin contribution. In addition, the bands display an alternating pattern of $S_z$ related by the $C_2$ symmetry. This behavior resembles that observed in the untwisted non-centrosymmetric bilayer, where the valence bands exhibit predominantly an alternating pattern of the out-of-plane spin polarization \cite{Jafari2024}. 

Finally, as indicated in Table \ref{tab:pst}, the PSTs lie along $\Gamma K$. This feature can be identified in Fig. \ref{fig:WSe2}c-f from the green arrows that are purely along the three stars of the $\Gamma-K$ line, with zero $S_z$. On the $\Gamma-M$ line, which is perpendicular to one of the $C_2$ axes, the $C_2\mathcal{T}$ symmetry maps $\vec{k}$ to itself and preserves only the perpendicular spin components $C_2\mathcal{T}\vec{S} = \vec{S}_{\perp}$. Therefore, even though $\Gamma-M$ is not a symmetry-enforced PST line since $S_z$ can also be finite, the in-plane component of the spin polarization is purely along the $\Gamma M$ direction.

Figure \ref{fig:WSe2}g shows the REE susceptibility calculated as a function of chemical potential. The results are consistent with predictions in Table \ref{tab:pst}, with $\chi_{xy}=\chi_{yx}=0$ and $\chi_{xx}=\chi_{yy}$. The values are approximately one order of magnitude smaller than those obtained for AgI. The largest peak, exceeding $10^{9}\frac{\hbar}{\text{A}\cdot \text{cm}}$, occurs between 0.2 and 0.4 eV below the VBM and may be associated with the spin splitting observed along the $K-M$ line in this energy range, although direct comparison is difficult due to the contributions from the entire BZ. The relatively small spin response may partly reflect the dominant out-of-plane spin polarization $S_z$, which does not contribute to the in-plane REE. The orbital REE, shown in Fig. \ref{fig:WSe2}h, exhibits the same sign and similar energy dependence as spin in the considered energy window, with slightly larger magnitudes. Thus, the orbital contribution again adds constructively to the spin contribution, enhancing the induced total angular momentum accumulation. 

\section*{Discussion}
In summary, we have found a general mechanism for the unconventional Rashba-Edelstein effect in chiral twisted van der Waals bilayers, providing a 2D realization of chirality-dependent spin and orbital responses in a lateral geometry, contrasting with more common out-of-plane configurations \cite{Felser2025}. Based on symmetry analysis, Slater-Koster calculations, and DFT studies of representative materials with small commensurate supercells, we show that these responses arise naturally in twisted bilayers and are not restricted to a particular material or precise twist angle. This extends the study of unconventional REE, previously explored in chiral bulk materials, such as Te, and specific graphene/TMD heterostructures, to general twisted van der Waals materials, potentially including several experimentally relevant systems. 

2D materials also facilitate electrostatic gating, through which the chemical potential can be tuned to pronounced features in the REE spectrum, potentially enhancing the response or allowing the spin and orbital contributions to be distinguished when they have opposite signs, as found in InBi. The unconventional REE is also related to the symmetry-protected PSTs along high-symmetry lines, which may lead to extended spin lifetimes, although establishing this connection quantitatively will require further calculations and experimental studies. Beyond spintronics, twisted van der Waals bilayers are interesting for orbitronics, combining chirality-dependent orbital responses with clean and controllable interfaces. Further exploration of these effects will benefit from high-throughput calculations and approaches based on the modern theory of orbital magnetization to fully assess the role of orbital degrees of freedom.

\section*{Methods}
\noindent\textbf{First-principles calculations}.\\ The density functional theory calculations were performed using the Quantum ESPRESSO package. The ion-electron interactions were described using fully relativistic pseudopotentials obtained from the PSLibrary database, and spin-orbit coupling was included self-consistently in all calculations except for the relaxations. The exchange-correlation was treated within generalized gradient approximation and Perdew-Burke-Ernzerhof (PBE) parametrization. For both systems, the kinetic energy cutoff was set to 70 Ry, and the charge density cutoff to 560 Ry. Brillouin zone sampling was performed using the Monkhorst-Pack scheme with a 11$\times$11$\times$1 k-point mesh for the scf calculation, followed by a nscf calculation with a 22$\times$22$\times$1 k-point mesh. Relaxation was performed considering the convergence parameters of $10^{-4}$ Ry/bohr and $10^{-5}$ Ry for force and energy, respectively. The obtained lattice parameters for the supercells were $a=b=9.610$ \AA \: and $a=b=8.663$ \AA \: for AgI and WSe$_2$, respectively. Van der Waals interactions were included using the DFT-D3 correction \cite{Giannozzi2009, Giannozzi2017}.

For post-processing, the PAOFLOW code was employed to calculate the electronic structures, spin textures, Fermi surfaces, and Rashba-Edelstein susceptibilities \cite{PAOFLOW1, PAOFLOW2, PAOFLOW3}. An ultradense $50\times50\times1$ k-point grid was used to generate the interpolated Hamiltonians. The REE susceptibilities were calculated based on the semiclassical Boltzmann transport approach within the constant relaxation time approximation. The details of implementation are specified in the previous studies \cite{Analogs2023, PAOFLOW3}. 

For the orbital REE, the spin operator was replaced by the orbital angular momentum operator $\hat{L}$. For relativistic calculations in Quantum ESPRESSO, the pseudo-atomic orbital (PAO) basis functions are in the $|j,m_j\rangle$ basis. $\hat{L}$ was obtained by subtracting the spin operator from the total angular momentum operator, i.e. $\hat{L}_\alpha=\hat{J}_\alpha-\hat{S}_\alpha$ for $\alpha=x,y,z$ defined in $|j,m_j\rangle$. For nonrelativistic calculations, and our Slater-Koster model, which are defined in $|\ell,m\rangle$ basis, $\hat{L}$ can be directly constructed using orbital angular momentum ladder operators, e.g. $\hat{L}_\alpha=-i\varepsilon_{\alpha\beta\gamma}\ket{p_\beta}\bra{p_\gamma}$ for $p$ orbitals. 
\vskip 0.5 em

\noindent\textbf{Slater-Koster Hamiltonian calculations.} \\The Slater-Koster Hamiltonians are constructed for a range of commensurate twist angles, given by the pair of indices ($m,n$), i.e. the supercell for top and bottom layer are \[\begin{bmatrix}
    n & m \\ -m & n
\end{bmatrix},\quad \begin{bmatrix}
    m & n \\ -n & m
\end{bmatrix}.\]
The size of moir\'e cell is $n^2+m^2$ and the twist angle $\theta_{\text{top}}-\theta_{\text{bot}}=\tan^{-1}((m^2-n^2)/(2mn))$. The untwisted unit cell has dimension of $1\times1$ in unit of \AA~and an interlayer distance of 2 \AA. Each unit cell contains 1 atomic site with $p_x$, $p_y$, $p_z$ orbital, with equal onsite energy set to 0 eV. The interactions are given by
\[t_{\alpha\beta}=\left[\delta_{\alpha\beta}V_{pp\pi}+\hat{R}_\alpha\hat{R}_\beta(V_{pp\sigma}-V_{pp\pi})\right]/R^2\]
where $R=|\vec{R}|$ and $\hat{R}=(R_x,R_y,R_z)/R$ are the bond length and bond direction. We choose $V_{pp\sigma}=1.0$ eV and $V_{pp\pi}=-0.3$ eV, and ignore bondings outside a cutoff of 1.5 \AA~for intralayer interactions and 2.5 \AA~for interlayer interactions. Atomic SOC $\lambda\vec{L}\cdot\vec{\sigma}$ is required to generate spin-momentum locking, with $\lambda$ chosen to be 0.1 eV. This is on a scale comparable to that of many atomic species. For example, Carrier \textit{et al.} estimated $\Delta_{SO}=\frac{3}{2}\lambda=$ 0.086, 0.386 and 0.815 eV for $p$ orbitals of S, Se and Te \cite{Pierre2004}. 

A real-space Hamiltonian $\mathcal{H}(\vec{R})$ is constructed from a $3\times3\times1$ array of unit cells, sufficiently large to accommodate all hopping terms within the cutoff. The construction is done with the help of PythTB package \cite{pythtb}. Then within PAOFLOW package, $\mathcal{H}(\vec{R})$ is converted to k-space $\mathcal{H}(\vec{k})$ with k-mesh of $3\times3\times1$, and interpolated to a dense mesh of $n_k\times n_k\times1$ for subsequent evaluation of the Rashba-Edelstein response. We determine $n_k$ by the size of moir\'e lattice, and we find convergence of the Rashba-Edelstein response using $n_k=3\bigg\lceil \sqrt{\frac{8000}{m^2+n^2}}\bigg \rceil$.

\begin{table}[!h]
    \centering
    \begin{tblr}{
        colspec = {cc|cc},
        hline{1,Z} = {1}{-}{solid},
        hline{1,Z} = {2}{-}{solid}}
        $m,~n$ & Angle ($^\circ$) &  $m,~n$ & Angle ($^\circ$)\\ \hline
        9, 8 & 6.73 & 4, 1 & -28.07 \\
        7, 6 & 8.80 & 7, 4 & 30.51 \\
        6, 5 & 10.39 & 10, 3 & -33.40 \\
        4, 3 & 16.26 & 2, 1 & 36.87 \\
        6, 1 & -18.92 & 11, 4 & -39.97 \\
        3, 2 & 22.62 & 8, 3 & -41.11 \\
        9, 2 & -25.06 & 5, 2 & -43.60 \\
    \end{tblr}
    \caption{Square moir\'e cell indices used in the Slater-Koster model and their twist angles. When twist angle is negative, its sign is flipped, as well as the Rashba-Edelstein response, so that all angles are within (0$^\circ$, 45$^\circ$).}
    \label{tab:indices}
\end{table}

\noindent \textbf{Symmetry analysis of twisted homobilayers.} \\ Starting from the monolayer $L$ invariant under group $\bm{G}$ ($RL=L$ for $R\in \bm{G}$), the top and bottom layer can be represented by $R_z(\theta)L+d\hat{z}$ and $R_z(-\theta)L-d\hat{z}=R_z^{-1}(\theta)L-d\hat{z}$. Suppose $\bm{G}$ contains inversion and mirror planes. Using the fact that inversion $\mathcal{I}$ and horizontal mirror $M_z$ commute with rotation $R_z(\theta)$, applying them on top layer give $R_z(\theta)L-d\hat{z}$, which does not map to itself nor the bottom layer. For vertical mirror plane $M_{\hat{p}}$ along an arbitrary in-plane axis $\hat{p}$, the relation $M_{\hat{p}}R_z(\theta)=R_z(-\theta)M_{\hat{p}}$ holds. Because $M_{\hat{p}}(R_z(\theta)L+d\hat{z})=R_z(-\theta)L+d\hat{z}$ for top layer, and similarly for bottom layer, any vertical mirror plane will not make the twisted bilayer invariant, and therefore does not survive the twist operation.

Next, we show that if $\bm{G}$ contains in-plane 2-fold axis, they are preserved in twisted bilayer. An in-plane 2-fold rotation along an arbitrary in-plane axis $\hat{p}$ has the relation $C_{2\hat{p}}R_z(\theta)=R_z(-\theta)C_{2\hat{p}}$, similar to $M_{\hat{p}}$. However, since $C_{2\hat{p}}$ changes $z$ to $-z$, $C_{2\hat{p}}(R_z(\pm\theta)L\pm d\hat{z})=R_z(\mp\theta)L\mp d\hat{z}$, in-plane $C_2$ exchanges the top and bottom layer and is a symmetry operation of the twisted bilayer.
\vskip 0.5 em

\noindent\textbf{Persistent spin textures in twisted homobilayers.} Whether PST is enforced by symmetry at a particular $\vec{k}$ vector can be deduced from the transformation of the spin matrix element, defined as $\langle\sigma_\alpha\rangle_{ij}=\langle\psi_i|\sigma_\alpha|\psi_j\rangle$ with $i,j$ labeling the Bloch states $\psi$, under symmetry operations \cite{Oguchi2009, Kilic2025}. For a symmetry operation $g$ belonging to the little group of $\vec{k}$, a spin along the direction $\alpha=x,y,z$ is allowed if \[\langle g\sigma_\alpha g^{-1}\rangle_{ij}=\sum_{k,k'}D_{ik}(g)D_{jk'}^*(g)\langle \sigma_\alpha\rangle_{kk'},\] where $D_{ij}(g)=\langle\psi_i|g|\psi_j\rangle$ is one of the irreducible corepresentations (coirrep) of $g$ that describes how the Bloch states $\psi$ transform under $g$. In general, symmetry-enforced PST can arise on high-symmetry planes, lines and points where symmetry allows the spin to be only in one direction. 

In chiral twisted homobilayer, due to absence of inversion symmetry, bands are nondegenerate at general k-points, and only degenerate at the time-reversal invariant momentum (TRIM). Considering high-symmetry lines along a 2-fold axis, as the bands are nondegenerate, the above condition reduces to \[\langle g\sigma_\alpha g^{-1}\rangle_i=\langle \sigma_\alpha\rangle_i.\] A general $C_2$ spin rotation along an arbitrary axis $\hat{n}$ is given by $g(C_{2\hat{n}})=-i\hat{n}\cdot\vec{\sigma}$. Since $g(C_{2\hat{n}})\vec{\sigma}g^{-1}(C_{2\hat{n}})=\vec{\sigma}$ for a spin $\vec{\sigma}$ parallel or anti-parallel to $\hat{n}$, but $-\vec{\sigma}$ for a spin perpendicular to it, spins along any $C_2$ axis $\hat{n}$ are symmetry-enforced to point along that axis. 

At some of the TRIM points, symmetry can also allow PSTs, with the requirement that there exists a symmetry operation (other than identity) whose coirrep is ($\pm1\times$) identity matrix, i.e. $\pm\mathbb{I}_{d\times d}$. In such cases, \[\langle g\sigma_\alpha g^{-1}\rangle_{ij}=\sum_{k,k'}(\pm\delta_{ik})(\pm\delta_{jk'})\langle \sigma_\alpha\rangle_{kk'}=\langle \sigma_\alpha\rangle_{ij},\]
imposing direct relationship between pairs of band indices $(i,j)$. For 2$_1$ screw axis with fractional translation $\vec{\tau}$, its coirrep at $\Gamma$ point is not $\pm\mathbb{I}_{d\times d}$ because $g^2(\{C_{2\hat{n}}|\tau\})=-\mathbb{I}_{2\times 2}$. However, at zone-boundary TRIM points, $g^2(\{C_{2\hat{n}}|\vec{\tau}\})=+\mathbb{I}_{2\times 2}$ if $\exp(-2i\vec{k}\cdot\vec{\tau})=-1$. Then, since $\langle g\sigma_\alpha g^{-1}\rangle_{ij}=\langle\sigma_\alpha\rangle_{ij}$ when $\vec{\sigma}$ is parallel or anti-parallel to the 2$_1$ screw axis, PST at these degenerate points are symmetry enforced. We consider layer group p42$_1$2 as an example. At the X point (0,1/2), there is PST along [0,1] direction because of $\{C_{2y}|1/2,1/2,0\}$ whose coirrep is $\pm\mathbb{I}_{2\times2}$. 

\section*{Data availability}
The data that support this study will be openly available at
DataverseNL after the paper is accepted. 

\section*{Acknowledgments}
We thank Berkay Kilic and Sergio Alvarruiz for the helpful discussions. S.P.P. acknowledges the hospitality at Zernike Institute for Advanced Materials at University of Groningen, where part of the research was performed. S.P.P. is a doctoral student in PCEIM at UNAM and has received a fellowship (No. 2056697) from Secretaría de Ciencias, Humanidades, Tecnología e Innovación (SECIHTI). J.S. acknowledges the funding through the ERC Consolidator Grant FERRERO (No. 101170522) and the Dutch Research Council (NWO) under the {ENW-M-1} project with file number OCENW.M.22.063 (Grant ID https://doi.org/10.61686/ZHLOR83497). R.M.C.V. acknowledges financial support from DGAPA-UNAM through PAPIIT project IN109124 and DGTIC-UNAM through LANCAD-UNAM-DGTIC-385. The calculations were carried out on the Dutch national e-infrastructure with the support of SURF Cooperative (EINF-17379) and on the Hábrók high performance computing cluster of the University of Groningen.

\section*{Author contributions}
S.P.P. performed first-principles calculations under the supervision of Z.H. Z.H. proposed the idea, performed symmetry analysis and calculations using the Slater-Koster Hamiltonians framework. S.P.P., Z.H. and J.S. analyzed the data and wrote the manuscript. R.M.C.V. and J.S. supervised and coordinated the project. 

\section*{Competing interests}
The authors declare no competing interests. 

\bibliographystyle{naturemag}
\bibliography{references}
\end{document}